\documentclass[sigconf,review=false,balance=false]{acmart}
\makeatletter
\renewcommand\@formatdoi[1]{\ignorespaces}
\makeatother

\setcopyright{acmcopyright}
\copyrightyear{2024}

\acmYear{2023}
\acmVolume{YYYY}
\acmNumber{X}
\acmDOI{XXXXXXX.XXXXXXX}
\acmISBN{}
\acmConference{IWSPA '24, June 21, 2024, Porto, Portugal}
\begin{document}

\title[PETs for AI-Enabled Systems]{Privacy-Enhancing Technologies for Artificial Intelligence-Enabled Systems}

\author{Liv d'Aliberti}
\orcid{0000-0002-3139-5960}
\affiliation{
  \institution{Leidos Inc.}
  \city{Arlington}
  \state{VA}
  \country{USA}
}
\email{olivia.daliberti@leidos.com}

\author{Evan Gronberg}
\orcid{0009-0008-6229-2797}
\affiliation{
  \institution{Leidos Inc.}
  \city{Riverside}
  \state{CA}
  \country{USA}
}
\email{evan.m.gronberg@leidos.com}

\author{Joseph Kovba}
\affiliation{
  \institution{Leidos Inc.}
  \city{Reston}
  \state{VA}
  \country{USA}
}
\email{joseph.g.kovba@leidos.com}

\renewcommand{\shortauthors}{d'Aliberti, Gronberg, Kovba}

\begin{abstract}
Artificial intelligence (AI) models introduce privacy vulnerabilities to systems. These vulnerabilities may impact model owners or system users; they exist during model development, deployment, and inference phases, and threats can be internal or external to the system. In this paper, we investigate potential threats and propose the use of several privacy-enhancing technologies (PETs) to defend AI-enabled systems. We then provide a framework for PETs evaluation for a AI-enabled systems and discuss the impact PETs may have on system-level variables.   
\end{abstract}


\maketitle

\section{Introduction}

Standard means of data protection provide strong security guarantees for data at rest and in transit. Encryption, access control, identity management, secure tunnels, firewalls, traffic monitoring, multi-factor authentication, and device management are all examples of current-generation practices that help ensure data remains protected and only accessible to its intended users. However, while all of these techniques fulfill their purpose of ensuring data is protected at rest and in transit, none of them address the protection of data in use.

To be used, data typically must be converted to its unprotected form (i.e., its plaintext). This applies equally to the use of data by humans and the use of data by machines. The plaintext must be accessible and available in both cases. Unfortunately, this creates an opportunity for data to be exposed to unauthorized parties, whether intentionally by malicious actors or unintentionally by careless users.

Due to the increase of real-world artificial intelligence (AI)-enabled systems, \cite{IBM2022} it is now more important than ever to address data-in-use concerns. This is because AI-enabled systems are uniquely reliant on data-in-use processes, whereas traditional systems typically rely on explicit, pre-programmed instructions to carry out tasks. The data used with AI-enabled systems both in training and in inference, as well as AI models themselves, are all involved in data-in-use processes. To address this concern, AI engineers are turning to next-generation protections for their systems: namely, privacy-enhancing technologies (PETs). 

\subsection{Data in Use for AI-Enabled Systems}

The term "data in use" refers to any data that is actively being processed (created, accessed, updated, or erased) by a human or a machine. In the context of AI, a human process could be performing data analysis; a machine process could be model training or performing model inference. 

Attacks against data-in-use processes at large have been well researched, \cite{Pitro19, Exposed2017} and these general attacks certainly pose a threat to AI-enabled systems. However, AI-enabled systems also have their own unique set of potential attack vectors, such as those described below.

\begin{itemize}

\item An AI model trained on private data could be used in ways that are unintended by the people who provided the data. A major component of privacy is control over how and when personal information is used. Technical guarantees need to be put in place to ensure that models trained on private data will only be used as permitted by the data owners.

\item An AI model trained on private data could be inappropriately shared and used to extract information about the data used to train the model. Protections need to be established to ensure that models will only be distributed and accessed in ways that are permitted by the data owners.

\item An AI model trained on private data could be manipulated in ways that may impact the behavior of the model. Models are highly refined, structured, and trained representations of data; both the way a model is structured and its training curriculum are key to its performance. Defenses need to be created to ensure the model, even in its intended system, is only executed as permitted by the data owners.

\end{itemize}

Ideally, AI models will only be run by model owners, as explicitly permitted by data owners, in a way that guarantees the data owner's privacy. This is difficult, because AI models have the potential to be attacked at any point in the model's lifecycle: data set creation, model development, model deployment, user access, or user execution. Protections need to be put in place at each of these stages to ensure the system does not violate data owner privacy.

\subsection{Privacy and Confidentiality}

Not all sensitive data is the same. Level of sensitivity and, consequently, level of protection required differ depending on specific use case and associated legal standards. These distinctions are important because they lead to different requirements for a system’s architecture. Two important, high-level concepts when dealing with sensitive data are privacy and confidentiality.

\begin{itemize}
\item {\bfseries Privacy} is control over the extent, timing, and circumstances of sharing personal information. Respecting privacy requires special protections around the ways personal information is collected, used, retained, disclosed, and destroyed. \cite{uci-privacy-confidentiality}

\item {\bfseries Confidentiality} is the protection of any information that an entity has disclosed in a relationship of trust with the expectation that it will not be divulged to unintended parties. \cite{edtech-confidentiality}

\end{itemize}

The primary difference between privacy and confidentiality is that the former concerns personal information and the latter concerns any data considered sensitive. Moreover, confidentiality concerns the unauthorized use of information already in the hands of an organization, \cite{nist-confidentiality} whereas privacy concerns the rights of an individual to control the information that an organization collects, uses, and shares with others. 

Of particular concern is Personally Identifiable Information (PII), "[a]ny representation of information that permits the identity of an individual to whom the information applies to be reasonably inferred by either direct or indirect means." \cite{dol-pii} PII data has legal, contractual, and ethical requirements that restrict its disclosure, and its use is controlled by data privacy regulations (e.g.,  HIPAA \cite{hipaa}, GDPR \cite{gdpr}).  
\subsection{Privacy-Enhancing Technologies (PETs)}

PETs are technologies that aim to protect data-in-use processes without preventing the system from accomplishing any of its necessary functions. Specifically, PETs are designed to do the following:

\begin{itemize}
\item Allow parties to collaborate while guaranteeing that any shared data will be used only for its intended purposes.
\item Glean insights from private data without revealing the sensitive contents of the data.
\item Carry out trusted computation in an untrusted environment.
\item Secure access to shared artificial intelligence (AI) models without revealing sensitive data.
\item Add quantum-resistant data protections to the system.
\item Enhance the ability of data owners to maintain control of their data throughout its lifecycle.
\end{itemize}

All of the aforementioned tasks are concerned with the protection of sensitive data and addressing data layer vulnerabilities. In practice, the term “privacy-enhancing technologies” encompasses a wide variety of tools – whether implemented via hardware or software, on premises or in the cloud – that are all designed to protect data-in-use processes. Figure 1 organizes various PETs by their primary characteristics and demonstrates the breadth of tools and functionalities that fall under the “PETs” umbrella.

\begin{figure}[h]
  \centering
  \includegraphics[width=\linewidth]{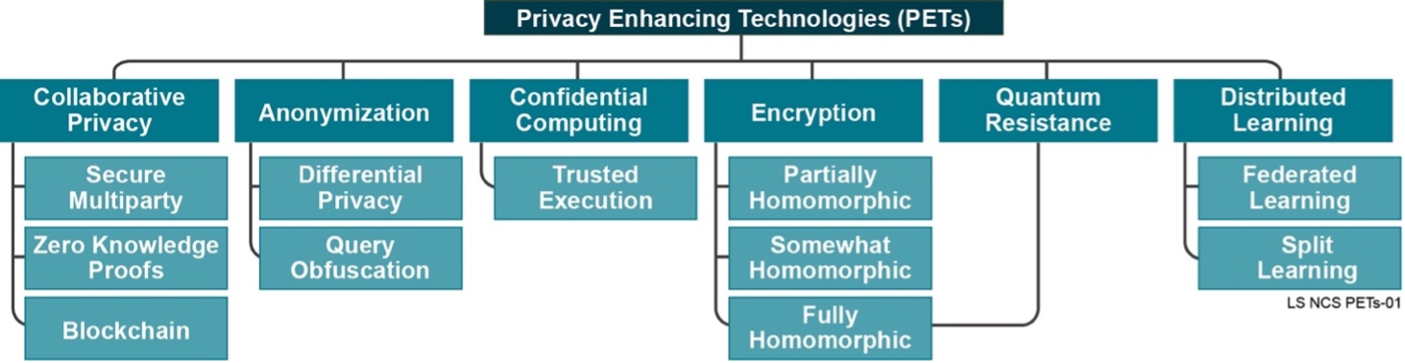}
  \caption{A selection of PETs organized into groups based on each technology’s primary characteristic.}
  \Description{Privacy-enhancing technologies broken down into subgroups of collaborative privacy, anonymization, confidential computing, encryption, quantum resistance, and distributed learning.}
\end{figure}

Multiple PETs may operate in tandem to address a particular security concern or set of security concerns more fully. As Figure 2 shows, each PET has its own strengths and weaknesses; there is no “one size fits all” PET.

\begin{figure}[h]
  \centering
  \includegraphics[width=\linewidth]{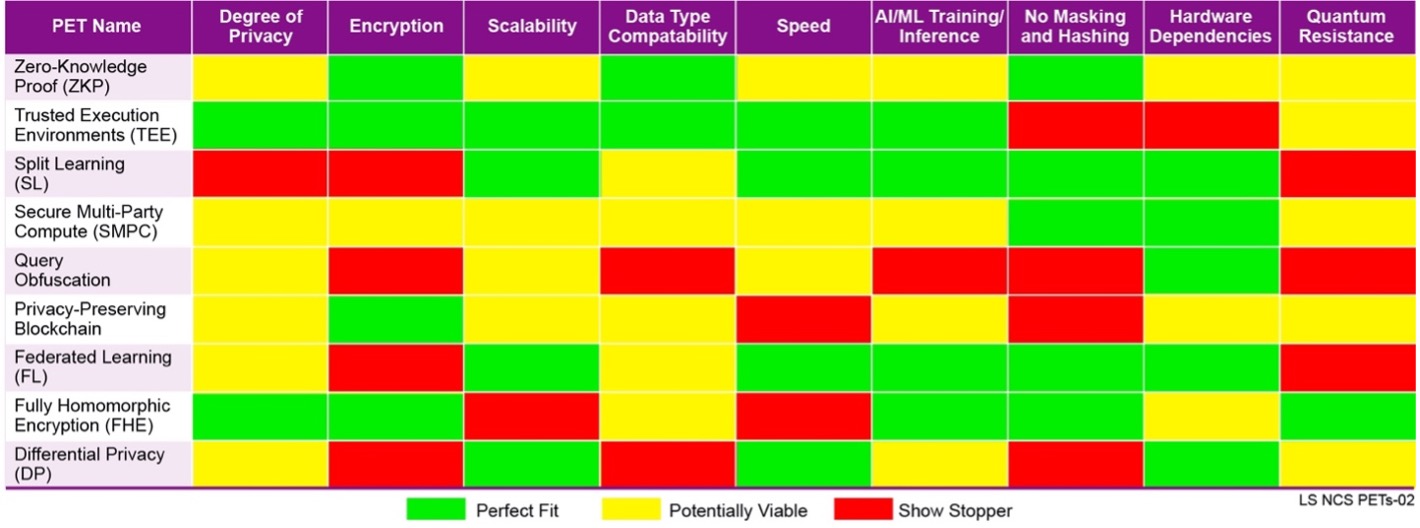}
  \caption{A selection of PETs evaluated against important system-level characteristics.}
  \Description{A stop-light diagram of different privacy enhancing technologies organized by their system-level impact upon implementation}
\end{figure}

The columns in the figure represent system-level characteristics that could be affected by the addition of a given PET. The primary capabilities of each PET are shown in this chart as a “Perfect Fit.” The “Potentially Viable” assignment means that, while the PET in its standard form could pose a problem, it can be modified to perform better in that category. The chart also shows why a given PET ought not to be used. For example, fully homomorphic encryption (FHE) can be very detrimental to a system’s speed and scalability, so it is classified as a "Show Stopper" for systems dependent upon those features.

PETs have great potential to improve the security of existing systems. Perhaps more notably, though, they have the potential to enable new systems that were previously not feasible due to data protection concerns. In particular, with PETs, new sources of data can be leveraged while still maintaining PII protections.

\section{Threats to AI-Enabled Systems}

AI-enabled systems require the protection of two high-level assets: the AI model itself and the data that is used in conjunction with that model. Opportunities for attacks against these assets exist throughout the system's lifecycle:

\begin{itemize}
    \item During development, researchers should be aware that certain model architectures are more vulnerable to privacy attacks, \cite{zhang2023does} and data sets can be manipulated to control the behavior of trained AI models, \cite{data-poisoning} thus delivering false or even harmful results.
    \item During deployment, access to model architecture, parameters, and/or ontology \cite{mitre} may result in the exfiltration of sensitive information. \cite{wsj-tiktok-attack}
    \item During inference, intentional misuse of data or a model could result in a re-linking \cite{record-linkage, netflix, gpt2extraction} or adversarial attack \cite{adversarial-attacks} and may result in privacy violations or unexpected model behavior.
\end{itemize}

The nature of a privacy-threatening attack will depend on the threat’s motivation and level of access to the system, as determined by their association with the organization that owns the system. Below we have identified three broad threat types:

\begin{itemize}
    \item Careless Insider: A person who has approved access to the system and has no malicious intent, but who nonetheless uses or leaks system information in unintended ways.
    \item Malicious Insider: A person who has approved access to the system and intends to abuse that access to harm the system or wrongfully distribute its contents.
    \item Outsider: A person who does not have approved access to the system, but who nonetheless intends to access it and harm the system or wrongfully distribute its contents.
\end{itemize}

Each of these threat types behaves uniquely and presents their own set of challenges. In the following sections, we elucidate potential attacks that each of them pose to AI-enabled systems.

\subsection{Insider Threats}

Insiders are responsible for 55\% of cybersecurity incidents. Of this 55\%, 57\% attack intentionally and 43\% act inadvertently. \cite{threats-breakdown} Insider incidents within non-AI-enabled systems are managed using practices like the principle of least privilege, user and entity behavior analytics (UEBA) \cite{ueba}, and general data loss prevention (DLP) practices \cite{dlp}. However, the rise of AI-enabled systems offers new ways for insider threats to manifest themselves in an organization.

For example, data leaks can occur when sensitive information is sent to a third-party AI model owner, as seen in Figure 3. A well-known example of a publicly available third-party model is OpenAI’s ChatGPT. \cite{chatgpt} As it stands, ChatGPT must have access to plaintext prompts to be able to produce responses. This requirement poses a confidentiality concern, since user input is necessarily exposed during ChatGPT’s inference process. Organizations that own models, like OpenAI, thus have direct access to the data sent by users.

\begin{figure}[h]
  \centering
  \includegraphics[width=\linewidth]{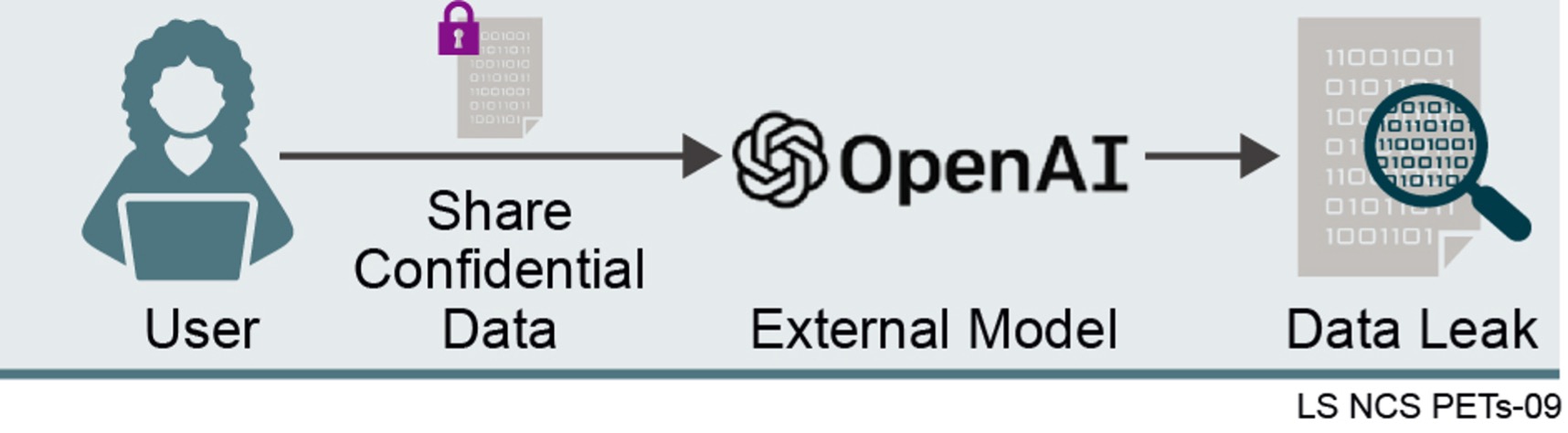}
  \caption{Example of an insider threat where a careless insider exposes potentially sensitive data to an external model.}
\end{figure}

The risk of careless insiders unintentionally revealing confidential data to AI-enabled services has led companies such as Samsung \cite{samsung} and Northrop Grumman \cite{northrop-grumman} to establish policies for the careful use of ChatGPT and other publicly available large language models (LLMs). However, the existence of policy does not provide guarantees that employees will follow procedure. Employees still have the ability to inadvertently expose sensitive data to external services without realizing the impact of their actions. 

Another unique challenge in the era of AI is that internal model developers now have access to an increased amount of data, some of which may be highly sensitive. For reference, an LLM like ChatGPT required a corpus size of 300 billion words, \cite{chatgpt-size} whereas previous-generation models, like BERT only required 3.3 billion words. \cite{bert-size} Access to big data opens up the potential for malicious insiders to exfiltrate and release large swaths of information. Careless insiders may also leak these data sets, albeit inadvertently, but it is the concentration of so much data in the hands of potentially malicious people that particularly demands our concern. Furthermore, while malicious insiders may also feed sensitive data to external services like ChatGPT, truly malicious actors will turn their attention toward much higher impact leaks such as model training sets and collected inference data.

Figure 4 shows how data, even data that is secured through standard, recommended security protocols, must normally be decrypted for any form of analysis, putting the data at risk of being distributed. Policy-based solutions, such as least access principles, are also often applied to this problem to minimize access to sensitive data, but few technological methods have been employed to ensure that malicious insiders do not have the ability to gather and distribute model data sets or inject poisoned data into a curated data set.

\begin{figure}[h]
  \centering
  \includegraphics[width=\linewidth]{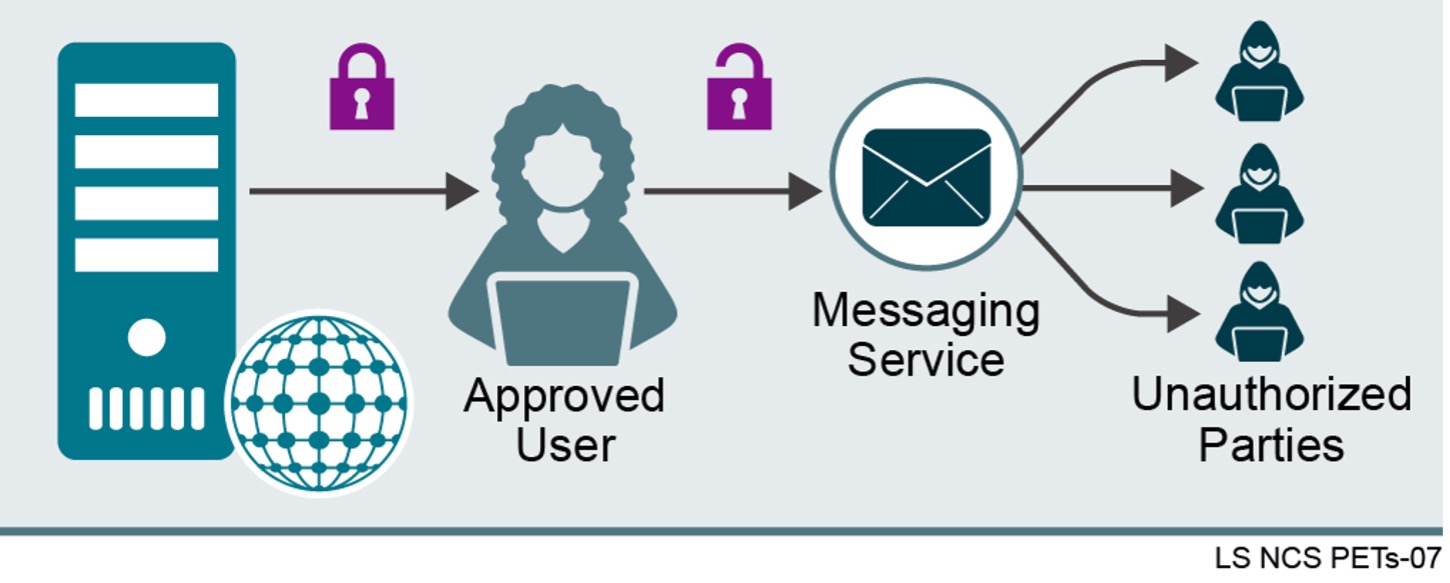}
  \caption{Example of an insider threat where a malicious insider exposes sensitive data to unauthorized parties via some messaging service.}
\end{figure}

\subsection{Outsider Threats}

Outsiders are responsible for the remaining 45\% of cybersecurity incidents. \cite{threats-breakdown} Outsider incidents with non-AI-enabled systems are managed using practices like intrusion detection and prevention systems (IDPS), standard encryption (i.e., in transit and at rest), firewalls, and robust authentication and access control. These work well to defend static data and maintain permissions boundaries. AI-enabled systems work differently though, because the data interaction is not static. Instead of connecting to deterministic end services that have been tightly defined, users are interacting directly with non-deterministic models. This interaction has the potential to be abused through reverse engineering, by untrusted third-party hosts, or as part of an adversarial attack.

A major privacy threat when deploying AI models is that models can potentially be reverse-engineered to reveal sensitive characteristics both about the model itself \cite{wsj-tiktok-attack} (which is a problem if the model is proprietary) and about the data that was used to train the model. Those characteristics may even enable someone to re-link the data that was used to create the model to PII data in the model’s training set. This scenario could result in an unintentional release of sensitive data, like in the Netflix Prize, a challenge in which researchers were able to take an anonymized data set and re-link people to their ratings of television shows. \cite{netflix} 

Also, training AI models can be a very computationally intensive process. The training of ChatGPT costs 3,640 PetaFLOP days. \cite{chatgpt-petaflops} Thus, compute is regularly outsourced to third parties. If we place sensitive data on third-party servers, we put a high amount of trust that the compute provider will not observe, collect, or tamper with data in unapproved ways.

A final threat posed by outsiders is an adversarial attack. Like any other data, AI models are vulnerable to manipulation, and adversarial attacks manipulate models by sending them examples that are specifically crafted to fool the model (see Figure 5). For instance, researchers highlighted the impact of adversarial attacks by confusing a Tesla Model S and forcing it to moving into oncoming traffic. \cite{tesla-attack}

\begin{figure}[h]
  \centering
  \includegraphics[width=\linewidth]{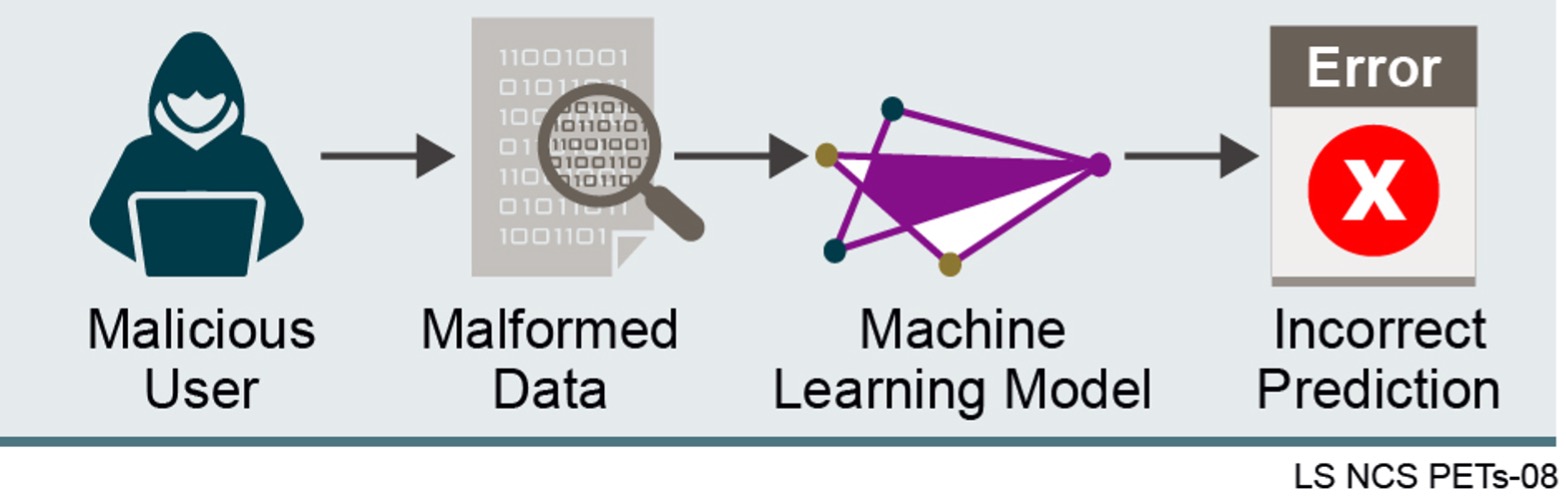}
  \caption{Example of an outsider threat where an outsider attempts to confuse an AI model by sending malformed data.}
\end{figure}

\subsection{Summary of Risks}

A typical, deployed AI-enabled system concerns two types of personnel, system owner and system user, and both are subject to privacy risks.

The owner of the system seeks to protect both the AI model and the data that was used to train it. Protecting the model means keeping its characteristics private, but it also includes maintaining the integrity of the model and its predictions against inputs purposefully engineered to fool or otherwise tamper with the model. Protecting the training data means preventing users from using the model to discover characteristics about the data. It also means preventing any data, whether with respect to the model or the training set, from being improperly leaked by those who develop and/or maintain the system.

The user of the system seeks to protect the data that is sent to and processed by the system. When a user requests an inference over private data from an AI-enabled system, they need to be assured that their personal data will not be exposed – and in many cases this means not having the data exposed even to the owner of the system. AI-enabled systems are often capable of learning the characteristics and behaviors of users over time, thus it is important that users be assured that their data will not be used for purposes they do not intend or desire.

PETs are effective tools for defending against system owner and system user vulnerabilities. Different PETs address different attacks from the different adversary types, and multiple PETs can be combined to either address multiple vulnerabilities or more robustly address a single vulnerability.

\section{PETs for AI-Enabled System Architectures}

PETs can be used to address the threats discussed above. The particular PETs listed below are some of the leading, most effective PETs for defending the vulnerabilities particular to AI-enabled systems. They are not the only PETs that be used for this purpose, but serve as great examples of the impact PETs can have toward next-generation security.

\subsection{Trusted Execution Environments (TEEs)}

A TEE is an isolated compute space, separate from its host instance, that relies on hardware-based encryption to protect both data in memory and application-level code. Entities outside a TEE, including the host instance, are unable to see or alter the code or data used during execution. A program running inside a TEE is cryptographically attested to be the program that is intended to run. \cite{tee-attestation} Figure 6 shows how a TEE could be deployed to ensure that data sent to a model is neither seen nor tampered with during inference.

\begin{figure}[h]
  \centering
  \includegraphics[width=\linewidth]{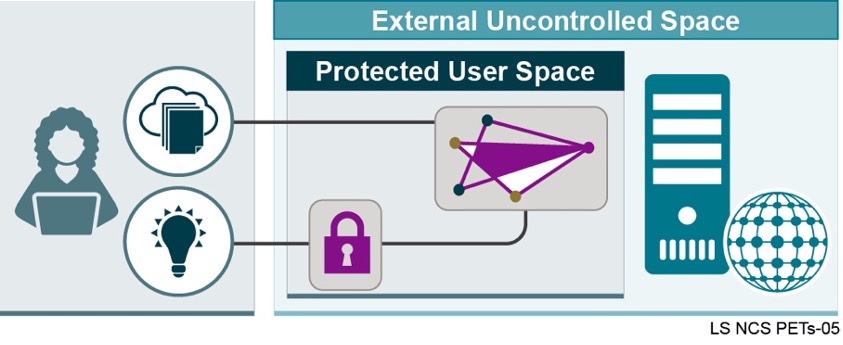}
  \caption{Example of a TEE Architecture for a Data Protection Scenario.}
  \Description{An architectural diagram of how TEEs might become part of a client-server architecture with an AI model in an external uncontrolled space}
\end{figure}

Using a TEE ensures that systems are protected from external compute providers, which is of great concern to system owners. Using a TEE also provides guarantees to system users, ensuring that their data is not logged or stored in ways outside of its intended use.

\subsection{Fully Homomorphic Encryption (FHE)}

FHE allows users to perform computation on encrypted data and models; that is, data in use is never decrypted. However, in its current-day state, it is very time- and space-intensive, so it must be implemented very conscientiously. Within the context of AI, a realistic use case for FHE is performing secure AI inference over smaller models in cases where increased inference latency is permissible. For example, DARPA is currently working on a program called Data Protection in Virtual Environments (DPRIVE) which aims to run inference over a 7-layer convolutional neural network performing inference against CIFAR-10 dataset in under 25 ms. \cite{DARPA} Figure 7 shows how FHE could be used to privatize an external, inference process.

\begin{figure}[h]
  \centering
  \includegraphics[width=\linewidth]{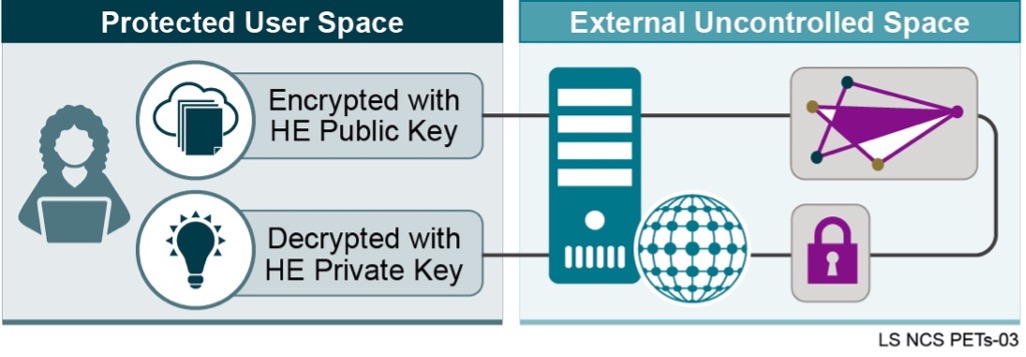}
  \caption{Example of an FHE Architecture for an Inference Data Protection Scenario.}
  \Description{An architectural diagram of how FHE might become part of a client-server architecture with an AI model in an external uncontrolled space}
\end{figure}

FHE allows encrypted interaction with an AI model and data sharing without risking misuse. There is strong potential that the time and space requirements of FHE will be improved through algorithmic \cite{CKKS-Network}, software, \cite{openfhe, BTS}
and hardware \cite{intel-fhe} developments. It is expected that upcoming developments in FHE will result in speed-ups of at least five orders of magnitude \cite{HERACLES}, and we expect to see further improvements as the technology continues to mature.

\subsection{Federated Learning (FL)}

Federated learning (FL) uses multiple clients’ data to build a shared AI model while keeping each client’s training data local; no client can access any other client’s data. Each client trains a model locally then sends that model (along with the number of examples that were used to train the model) to the server. The server then performs a weighted aggregation of all the models and sends the aggregated model to each client in the federation. This architecture relies on edge nodes with enough compute capacity to update their local model and may risk data de-anonymization \cite{fl-gla-1, fl-gla-2} if not used in tandem with another PET. Figure 8 shows an example of how FL could be used in a centralized system to update client models without sharing client data.

\begin{figure}[h]
  \centering
  \includegraphics[width=\linewidth]{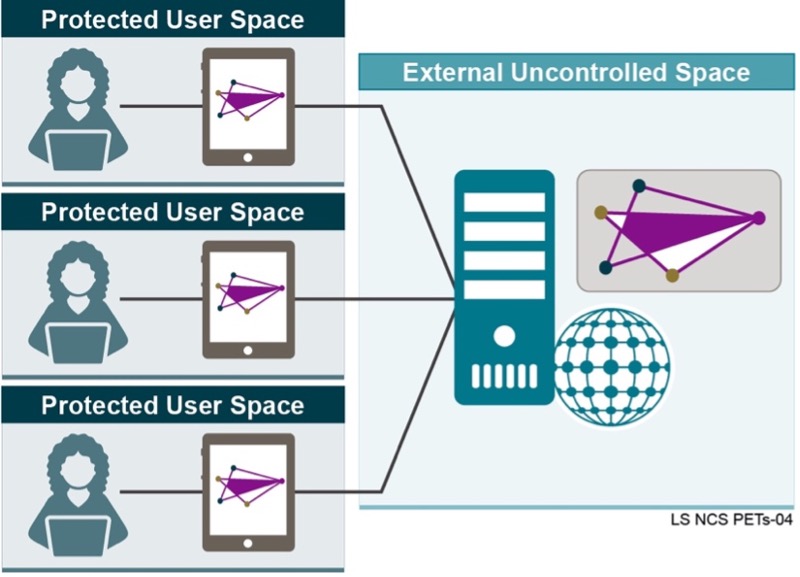}
  \caption{Example of an FL Architecture for a Data Protection Scenario.}
  \Description{An architectural diagram of how FL might become part of a client-server architecture with an AI model in an external uncontrolled space}
\end{figure}

FL directly addresses the concern of malicious insiders within the context of a collaborative group. Normally, distinct parties would be very hesitant to work together on AI models, because then all parties' data needs to be shared. One party could reveal the sensitive contents of others' data, even if that party had previously been deemed an "insider." FL greatly reduces the amount of trust required to collaborate on AI models.

With FL, the collaborators typically use a third-party server to perform the aggregation of their models. They can leverage an outsider for this because only model weights are sent to the server, not any data. However, to prevent the server from reverse-engineering these model weights, another PET such as FHE can be incorporated.

\subsection{Summary of PETs Selection}

The three PETs in this section have been selected for both their high utility and their wide applicability to a range of AI-related use cases. TEEs and FHE are both especially useful for securing the inference process. \cite{tee-inference, fhe-inference} With them, an organization can make an AI model available to a group people without ever seeing the data those people send to the model. FL is particularly useful for the training process; data from many parties can be leveraged without being revealed. TEEs and FHE can also be used for the training process, although the time and space demands of FHE in particular make training very difficult to scale. \cite{fhe-speed} The use of TEEs for training is more feasible than the use of FHE, but (1) it does not provide the same distributed computing benefit that FL does and (2) current, leading implementations of TEEs only isolate CPU and RAM (that is, they cannot leverage GPUs, and they do not have non-RAM storage, so they have very high memory requirements). \cite{aws-nitro-enclave, azure-sgx-enclave}

These three PETs were thus selected because together they cover the main phases of the AI-enabled systems lifecycle, development/training and deployment/inference. There are, however, other PETs that can also be useful to AI-enabled systems:

\begin{itemize}
    \item Synthetic Data Generation: Data generated by humans is not used for training. Instead, "fake" data that is statistically similar to human data is used. \cite{synthetic-data-generation}
    \item Differential Privacy: Statistical "noise" is added to training data in order to obfuscate its finer, potentially sensitive, details. \cite{differential-privacy}
    \item Split Learning: The overall AI model is split such that the first portion of the model (typically a small portion) is placed on clients and the second portion of the model (typically a large portion) is placed on a server. The model is trained by sending client portion outputs to the server portion, which calculates the model updates. Thus raw client data is never shared. \cite{split-learning} 
\end{itemize}

These PETs were not described in detail because they pose more serious tradeoffs compared to TEEs, FHE, and FL, particularly in regards to model performance. They are, however, still worth consideration if implemented conscientiously.

\subsection{PETs for Model Integrity}

The above PETs address the confidentiality of data and models, but they do not address the integrity of model behavior. In the same way that current-generation cryptographic methods are used to ensure the integrity of data at rest and in transit (e.g., hashing), next-generation PET protections ought to address the integrity of AI models. The goal of hashing is to validate that data will accomplish its intended purpose, and it is important to carry this parallel into AI-related data-in-use scenarios.

As mentioned in section 2.2, adversarial attacks are a great example of a threat to the integrity of model behavior. Existing methods for addressing this threat include randomizing inference input data, \cite{adversarial-input-randomization}, purposefully including adversarial examples in training data, \cite{adversarial-training} regularizing models to prevent input data sensitivity, \cite{adversarial-regularization} or simply attempting to detect adversarial inputs directly. \cite{adversarial-detection} However, none of these methods provide formal guarantees that the model will behave as expected, and such guarantees are important for high risk models such as those used by autonomous vehicles. This, then, is an important gap that future PETs could help fill.

\section{Evaluating PETs}

Turning our attention to PETs as a whole, it is important to have a way to evaluate the fitness of any given PET for any given AI-enabled system. Although they all increase privacy, not all PETs perform the same function or provide the same level of protection and performance. Previously, in Figure 2, we listed many system-level characteristics to consider when designing a PETs solution for data layer security. We now outline a holistic framework for determining PET fitness.

\subsection{Use Case Applicability}

The first step of integrating a PET into an AI-enabled system is to clearly understand the system’s use case and formally define the utility that the PET would provide. The technical objective, the threats to the system, and all relevant privacy and security requirements should be documented. AI system architects with experience using PETs will be able to select an appropriate PET (or set of PETs) to protect against data-in-use vulnerabilities. To leverage this experience, many organizations are collaborating to collect, curate, and publish use case repositories for open consumption. \cite{enisa} Consider the following questions to guide a use case applicability assessment:

\begin{itemize}
    \item Does the given PET adequately address security requirements while maintaining the necessary level of AI model performance?
    \item Does the use case include specific privacy or confidentiality requirements?
    \item With what chief mechanism does the given PET operate: encryption, obfuscation, anonymization, or another mechanism?
\end{itemize} 

\subsection{System Impact}

PETs provide innovative protections, but they may also require performance tradeoffs. A clear set of standards for PET performance characteristics is still under development, although significant progress has been made in the past five years. \cite{fhe-standard, iso-fhe-standard} These up-and-coming standards aim to define cryptographic schemes and security parameters. Furthermore, standards for security and privacy best practices for AI-enabled systems in general are actively being developed. \cite{iso-ai} A key step toward implementing PETs in these systems is understanding new standards and how they relate to existing system requirements.

From an integration perspective, one must consider how the introduction of PETs affects the current system security level. PETs are still in their relative infancy, and new attacks against them will likely be developed in the future. PETs must be integrated in a way that allows for cryptographic agility; otherwise, they may have a net-negative impact on system security in the long run. However, with thoughtful design and a flexible integration plan, PETs have the potential to greatly enhance overall system security.

Currently, privacy and performance are typically at odds with each other; enhanced privacy can come at the cost of downgraded performance, and vice versa. PETs have the potential to undermine intended system functionality, and it is important to consider that tradeoff during evaluation. Consider the U.S. Census Bureau’s effort to minimize disclosure risk via differential privacy (DP). \cite{us-census} Though DP is effective at protecting the identities of individuals represented in a dataset, the “noise” that it injects into data necessarily perturbs it and can reduce statistical accuracy. The Census Bureau determines voting districts and allocates federal funding based on population statistics. Small perturbations in census data have the potential to produce outsized impacts on real-world decisions. Thus, in this case, statistical safeguards that protect the fidelity of data must be put in place to ensure the appropriate balance of privacy and accuracy. Another example might be a healthcare use case in which a doctor needs immediate access to records from another hospital as part of patient care. If the use of a PET introduces increased latency into a system, it could be detrimental to patient outcomes. Thus, it is critically important to be mindful of how the performance characteristics of a PET effect a given use case. The following questions are helpful in evaluating PET-related tradeoffs.

\begin{itemize}
    \item Does the type of security provided by the PET integrate well with existing, standard security measures already provided by the system?
    \item How flexible is the system in its ability to trade time, space, compute capacity, and other performance-related characteristic for added data protections?
    \item Where is the system hosted? On premises, in the cloud, or across both?
    \item Who controls the data collection process?
    \item Is the system centralized or decentralized?
\end{itemize}

\subsection{Implementation Readiness}

Implementation readiness concerns both the readiness of the technology itself and the readiness of an organization to adopt the technology. PETs are currently in a transitional state as they go from being prototyped in a lab to being deployed in the real world. Organizations pushing PETs toward deployment include large companies—such as Intel Corporation, a leader in TEEs \cite{intel-sgx}, specialized FHE hardware \cite{intel-fhe}, and remote attestation \cite{intel-remote-attestation}—and small companies—such as Duality \cite{duality}, a vendor with mature service offerings based on contributions to open-source FHE. \cite{duality-openfhe} While typical measures for software acceptance (e.g., static and dynamic code analysis, fuzzing, penetration testing) are important for guaranteeing the security of such systems, their cryptographic nature also requires formal guarantees of their functional characteristics. Organizations adopting PETs should implement formal verification methods to ensure that the system operates as expected. To aid in such efforts, some organizations are publicly publishing PET maturity assessments to reduce the level of effort required to fully assess a PET’s readiness level. \cite{enisa} An organization must also carefully consider its own readiness for adopting PETs. Though PETs focus on data security and privacy, they have wide-ranging implications for enterprise data architecture, data governance policies, potentially multinational legal requirements, and ethical matters (which greatly affect customer confidence). Moreover, an organization must consider standard technical matters, such as staffing needs (at the development, integration, deployment, and maintenance levels), security analysis and accreditation, and technological strategy around data and AI/ML. These two groups of factors, organizational and technical, must be evaluated jointly to ensure the most effective adoption of PETs. The following questions could help guide a discussion about adoption readiness:

\begin{itemize}
    \item Has the underlying mechanism been deployed in other real-world scenarios?
    \item Does the PET provider have a proven record of deploying software?
    \item How mature is the PET?
\end{itemize}

\section{Conclusion}

PETs represent the next step forward for cybersecurity. No longer can we only protect data at rest and in transit. We must also protect data in use, thus closing the final gap and providing true end-to-end security. Here we have highlighted how relevant PETs are to another technological leap: artificial intelligence. If the rise of AI-enabled systems is accompanied by the rise of PET-based protection, then the privacy risks associated with these systems can be greatly diminished. Overall, PETs directly pursue one of the foremost ethical principles of technology: "do no harm." In the era of AI, we believe that PETs are especially relevant to that goal.

\bibliographystyle{ACM-Reference-Format}
\bibliography{references}

\end{document}